\documentclass[wcp]{jmlr}

\jmlrvolume{1}
\jmlryear{2018}
\jmlrworkshop{Submitted to ICML 2018 AutoML Workshop}

\definecolor{lgreen}{rgb}{0.1,0.6,0.1}

\def\RSet{\mathbb{R}}

\usepackage{xcolor}

\usepackage{makecell}
\usepackage{capt-of}

\title[\textsc{Far-HO}]{\textsc{Far-HO}: A Bilevel Programming Package for Hyperparameter
Optimization and Meta-Learning}

  \author{\Name{Luca Franceschi} 
  \Email{luca.franceschi@iit.it} \\
    \addr Istituto Italiano di Tecnologia, Genova, Italy \& University College London, London, UK
  \AND
   \Name{Riccardo Grazzi} 
   \Email{riccardo.grazzi@iit.it} \\
   \addr Istituto Italiano di Tecnologia, Genova, Italy
   \AND
   \Name{Massimiliano Pontil} \Email{m.pontil@ucl.ac.uk} \\
   \addr Istituto Italiano di Tecnologia, Genova, Italy \& University College London, London, UK
   \AND
   \Name{Saverio Salzo}
   \Email{saverio.salzo@iit.it} \\
   \addr Istituto Italiano di Tecnologia, Genova, Italy
   \AND
   \Name{Paolo Frasconi} 
   \Email{paolo.frasconi@unifi.it} \\
   \addr{Universit\`a degli studi di Firenze, Firenze, Italy}
   }

\begin{document}

\maketitle

\begin{abstract}

  In \citep{franceschi_bilevel_2018} we proposed a unified mathematical framework, 
  grounded on bilevel programming, that encompasses gradient-based hyperparameter 
  optimization and meta-learning. We formulated an approximate version of the problem
  where the inner objective is solved iteratively, and gave sufficient conditions
  ensuring convergence to the exact problem. In this work we show how to optimize 
  learning rates, automatically weight the loss of single examples  and learn hyper-representations with \textsc{Far-HO}, a software 
  package based on the popular deep learning framework TensorFlow
  that allows to seamlessly tackle both HO and ML problems.\footnote{This submission
  is a reduced version of \cite{franceschi_bilevel_2018}
  which has been accepted at the main ICML 2018 conference. 
  In this paper we illustrate
  the software framework, material that could not be included 
  in the conference paper.}
\end{abstract}

\section{Introduction}

While in the standard supervised learning problems we seek the best
hypothesis in a given space and with a given learning algorithm, in
hyperparameter optimization (HO) and meta-learning (ML) we seek a
configuration so that the
optimized learning algorithm will produce a model that generalizes
well to new data.
The search space in ML often incorporates choices associated with the
hypothesis space and the features of the learning algorithm itself
(e.g., how optimization of the training loss is performed).  Under
this common perspective, both HO and ML essentially boil down to
\emph{nesting two search problems}: at the inner level we seek a
good hypothesis as in standard supervised learning while at the
outer level we seek a good configuration (including a good hypothesis
space) where the inner search takes place.  

HO and ML only differ substantially in terms of the experimental
settings in which they are evaluated. While in HO the available data
is associated with a single task and split\footnote{
Data could also be split multiple times, following a cross-validation scheme.}
into a training set, (used
to tune the parameters) and a validation set, (used to tune the
hyperparameters) in ML we are often interested in the so-called
\textit{few-shot}
learning setting where data comes in the form of short episodes 
sampled from a common
probability distribution over supervised tasks.
Algorithmic techniques for solving HO and ML can also differ
substantially. Classic approaches to HO, \citep[see e.g. ][and references therein]{hutter_beyond_2015} are only
capable of handling up to a few hundred hyperparameters.  Recent
gradient-based techniques for HO, however, have significantly
increased the number of hyperparameters that can be optimized
\citep{domke_generic_2012,maclaurin_gradient-based_2015,%
  pedregosa2016hyperparameter,franceschi_forward_2017} making it
possible to handle more hyperparameters than parameters.

As shown in \citep{franceschi_bilevel_2018}, HO and ML can be unified
within
the natural mathematical framework of bilevel programming, where an outer
optimization problem is solved subject to the optimality of an inner
optimization problem. 
The  variables of the outer objective are either the hyperparameters of a supervised learning problem in HO or the parameters of a meta-learner in ML.
In HO the inner problem is usually the minimization of an empirical loss, while in ML it could concern classifiers for individual tasks. Table \ref{tab:namings} outlines the links among bilevel programming, HO and ML. 

Bilevel programming~\citep{bard_practical_2013} has been suggested before in
machine learning \citep{keerthi2007efficient,kunapuli_classification_2008, flamary2014learning,pedregosa2016hyperparameter},
but never in the context of ML\@. The resulting framework, outlined 
in Section~\ref{sec:framework}, encompasses some existing approaches to
ML.
A technical difficulty arises when the solution to the inner problem
cannot be written analytically 
and one needs to resort to
iterative optimization approaches. We outline this approach in Section \ref{sec:gba} and we briefly discuss 
conditions that 
guarantee good approximation properties.

We developed a software package. \textsc{Far-HO}, based on the popular deep learning framework TensorFlow \citep{tensorflow2015-whitepaper} to facilitate the formalization and the solution of problems arising in HO and ML in a unified manner. We present an overview of the package and showcase two possible applications in Section \ref{sec:farho}.


\begin{table}[ht]
\begin{minipage}[c]{0.45\linewidth}
\centering
\caption{Links and naming conventions among different fields}
\label{tab:namings}
\begin{tabular}{c|c|c}
         \thead{Bilevel\\programming} &  \thead{Hyperparameter\\optimization} & \thead{Meta-learning} \\
         \hline \hline
         \thead{Inner variables} & \thead{Parameters} & \thead{Ground models'\\parameters} \\
         \thead{Outer variables} & \thead{Hyperparameters} & \thead{Meta-learner's\\parameters} \\
         \thead{Inner objective} & \thead{Training error} & \thead{tasks\\training errors} \\
         \thead{Outer objective} & \thead{Validation error} & \thead{Meta-training\\ error} \\
         \end{tabular}
    \label{table:student}
\end{minipage}\hfill
\begin{minipage}[c]{0.45\linewidth}
\centering
    \includegraphics[width=.98\textwidth]{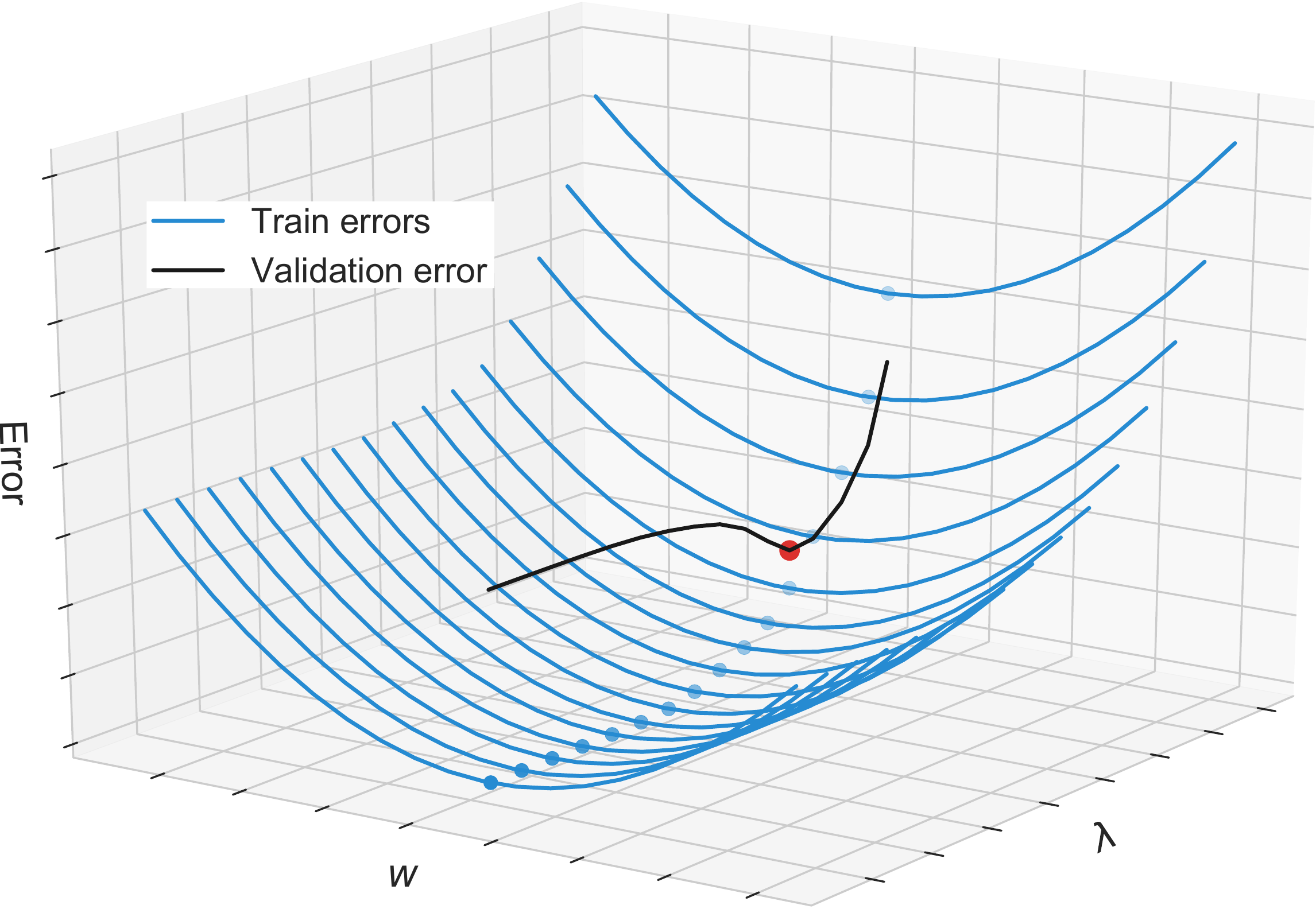}
\captionof{figure}{{\small Blue lines represent (average) training errors, varying $w$. The blue dots are the corresponding (inner) minimizers. The black line represents the outer objective $E$, whose minimizer is the red dot.
    }}
\label{fig:bilevel}
\end{minipage}
\end{table}

\section{A bilevel optimization framework}
\label{sec:framework}
We consider bilevel optimization problems \citep[see
e.g.][]{colson2007overview} of the form 
\begin{equation}
\min \{ f(\lambda) : \lambda \in \Lambda\},
\ \ \ \ \mbox{where} \ \ \ 
f(\lambda) = \inf \{ E(w, \lambda ) : w \in  {\rm arg}\min_{u\in \mathbb{R}^d} L_{\lambda}(u) \}.
\label{eq:f}
\end{equation}
Specific instances of this problem include HO and ML, which we discuss next.
We call $E:\mathbb{R}^d \times \Lambda\to\RSet$ the \emph{outer objective}, interpreted as a meta-train or validation error 
and, for every $\lambda \in \Lambda$, we call $L_{\lambda}:\RSet^d\to\RSet$ the \emph{inner objective}. 
$\{L_{\lambda} : \lambda \in \Lambda\}$ is regarded as
a class of objective functions parameterized by $\lambda$, wherein each single function may represent a training  error, possibly averaged multiple tasks. 
The cartoon in Figure \ref{fig:bilevel} depicts a stereotypical scenario.  


\subsection{Hyperparameter Optimization}

In the context of hyperparameter optimization, we are interested in
minimizing the validation error of a model
$g_w:\mathcal{X}\to\mathcal{Y}$ parameterized by a vector $w$, with
respect to a vector of real-valued hyperparameters $\lambda$.
For example, we may consider representation or regularization hyperparameters 
that control the hypothesis space or penalties, respectively.

In this setting, a prototypical choice for the inner objective is the regularized empirical error 
\begin{equation}
L_\lambda(w) = \sum_{(x,y) \in D_{\rm tr}}\ell(g_w(x),y) + \Omega_\lambda(w)
\end{equation}
where $D_{\operatorname{tr}}=\{(x_i,y_i)\}_{i=1}^n$ is a set of
input/output points,
$\ell$ is a prescribed loss function, and $\Omega_\lambda$ a regularizer
parameterized by $\lambda$. 

The outer objective represents a proxy for
the generalization error of $g_{w}$, and it is given by the average loss on a
validation set 
$D_{\operatorname{val}}$
\begin{equation}
E(w,\lambda) = \sum_{(x,y) \in D_{\rm val}} \ell(g_w(x),y).
\end{equation}
Note that the outer objective $E$ does not depend explicitly 
on the hyperparameters $\lambda$, since in HO $\lambda$ is instrumental in finding 
a good model $g_w$, which is our final goal.

\subsection{Meta-Learning}

In meta-learning (ML) the inner and outer
objectives are computed by respectively summing and averaging the training and the
validation error of multiple tasks. 
The goal is to produce a learning algorithm that will work well on novel 
tasks\footnote{This is also related to multitask learning, 
except in ML the goal is to extrapolate to novel tasks.}.

For this purpose, we have available a meta-training set
$\mathcal{D}=\{D^j=D^j_{\operatorname{tr}}\cup D^j_{\operatorname{val}}\}_{j=1}^{N}$, which is a
collection of datasets, 
sampled from a meta-distribution $\mathcal{P}$.
Each dataset
$D^j=\{(x_i^j,y_i^j)\}_{i=1}^{n_j}$ with
$(x_i^j,y_i^j)\in\mathcal{X}\times\mathcal{Y}^j$ is linked
to a specific task. Note that the output space is task dependent 
(e.g. a multi-class classification problem with variable number of classes). 
The model for each task is a function $g_{w^j,\lambda}:\mathcal{X}\to\mathcal{Y}^j$,
identified by a parameter vectors $w^j$ and hyperparameters $\lambda$. A key point
here is that $\lambda$ is shared between the tasks. With this notation the inner
and outer objectives are
\begin{equation}
  \label{eq:mlLE}
  \begin{array}{ccc}
    L_\lambda(w) = \sum_{j=1}^N L(w^j, \lambda, D^j_{\operatorname{tr}}),
    & \mbox{and}
    & E(w,\lambda) = \dfrac{1}{N} \sum_{j=1}^N L(w^j, \lambda, D^j_{\operatorname{val}})
  \end{array}
\end{equation}
respectively, where $L(w,\lambda,S)$ is the empirical loss of the pair
$(w,\lambda)$ on a set of examples $S$.
%
Particular examples are choosing  
the model $g_{w,\lambda} = \langle w, h_\lambda(x)\rangle$, 
in which case $\lambda$ parameterizes a feature mapping, or
consider
$g_{w^j,\lambda}(x) = \langle w + \lambda, x \rangle$, in which case $\lambda$ represents a common model around which task specific models are  to be found.

Note that the inner and outer
losses for task $j$ use different train/validation splits of the
corresponding dataset $D^j$. Unlike in HO, in ML the final goal is to
find a good $\lambda$ and the $w^j$ are now instrumental.

\subsection{Gradient-Based Approach}  \label{sec:gba}
We now discuss a general approach to solve Problem (\ref{eq:f}).
%
In general there is no closed form expression $w_\lambda$, so it is not possible to directly optimize the
outer objective function. A compelling approach is to replace the 
inner problem with a dynamical system, as discussed in \citep{domke_generic_2012,maclaurin_gradient-based_2015,franceschi_forward_2017}.

Specifically, we let $[T]=\{1,\dots,T\}$ where $T$ is a prescribed positive integer and consider the following approximation of Problem (\ref{eq:f})
\begin{equation}
\label{eq:general:constrained}
\min\limits_{\lambda} f_T(\lambda) = E(w_{T,\lambda}, \lambda), \hspace{6mm} \mbox{s.t.} \hspace{6mm}
w_{0,\lambda} = \Phi_0(\lambda),~w_{t,\lambda} =  \Phi_t(w_{t-1,\lambda},\lambda),~t \in [T],
\end{equation}
with $\Phi_0:\RSet^m\to\RSet^d$ a smooth initialization mapping and,
for every $t \in [T]$,
$\Phi_t : \RSet^d \times \RSet^m \rightarrow \RSet^d$ a smooth mapping
that represents the operation performed by the $t$-th step of an
optimization algorithm such as gradient descent\footnote{Other
  algorithms such as Adam requires auxiliary variables that need to be
  included in $w$.}, i.e.
$\Phi_t(w_t) = w_t - \eta_t \nabla L_{\lambda}(\cdot)$.

The approximation of the bilevel problem~\eqref{eq:f}
by Procedure~\eqref{eq:general:constrained} raises the issue of the
quality of this approximation:
$f_T$ may, in general, be unrelated to $f$. Among the possible issues, we note that, for a chosen $T$, $w_{T, \lambda}$ may not even be an approximate minimizer of $L_{\lambda}$, or, in the presence of multiple minimizers, the optimization dynamics may lead to a minimizer which not necessarily achieves the infimum of $E$. The situation is, however, different if the inner problem admits a unique minimizer for every $\lambda\in\Lambda$ (e.g. when $L_{\lambda}$ is strongly convex). In this case, it is possible to show, under some regularity assumptions, that the set of minimizers of $f_T$ converges to that of $f$ for $T\to\infty$ \citep{franceschi_bilevel_2018} and that the gradient of $f_T$ converges to $\nabla f$.

On the other hand, Procedure \eqref{eq:general:constrained} also suggests to consider the inner dynamics as a
form of approximate empirical error minimization 
which is valid in its own right. From
this perspective it is possible (and indeed natural) to include among
the components of $\lambda$ variables associated with the
optimization algorithm itself. 
For example, in \citep{de_freitas_learning_2016,wichrowska2017learnedICML} the
mapping $\Phi_t$ is implemented as a recurrent neural network, while
\citep{finn_model-agnostic_2017} focus on the initialization mapping
by letting $\Phi_0(\lambda) = \lambda$.

A major advantage of the reformulation above is that it makes it possible to compute efficiently  
the gradient of $f_T$, called \emph{hypergradient},
either in time or in memory
\citep{maclaurin_gradient-based_2015, franceschi_forward_2017}, by
making use of Reverse or Forward mode algorithmic differentiation 
\citep{griewank2008evaluating}. 
This makes it 
feasible to efficiently search for a good configuration in a high dimensional hyperparameter space -- reverse mode having a complexity in time independent form the 
size  
of $\lambda$.





\section{\textsc{Far-HO}: A Gradient-Based Hyperparameter Optimization Package} \label{sec:farho}

We developed a software package in Python with the aim to facilitate the formalization, implementation and numerical solution of HO and ML problems with continuous hyperparameters
(e.g. ridge regression, logistic regression, deep neural networks, ...). \textsc{Far-HO}, available at \url{https://github.com/lucfra/FAR-HO},  
is based on the popular library TensorFlow~\citep{tensorflow2015-whitepaper}; 
by leveraging the computational power of modern GPUs, it allows to scale up gradient-based
hyperparameter optimization techniques to high dimensional problems (many parameters and/or
hyperparameters).  Notably, it implements dynamic forward and reverse mode iterative differentiation\footnote{
It is not required to ``unroll'' the computational graph of the optimization dynamics 
for $T$ steps. Unrolling quickly becomes impractical as $T$ grows. 
}
for Procedure~\ref{eq:general:constrained} and warm restart strategies 
for the optimization of the inner problem and the computation of the hypergradient. 
The package exposes utility functions to declare hyperparameters, instantiate commonly used first-order optimization dynamics such as gradient descent or Adam and features single-line calls to set up approximate bilevel problems. Two practical examples are illustrated in the
remainder of this section. Experimental results obtained with
\textsc{Far-HO} are reported in~\citep{franceschi_bilevel_2018}.

\begin{figure}[h!]
\centering
\includegraphics[width=0.8\textwidth]{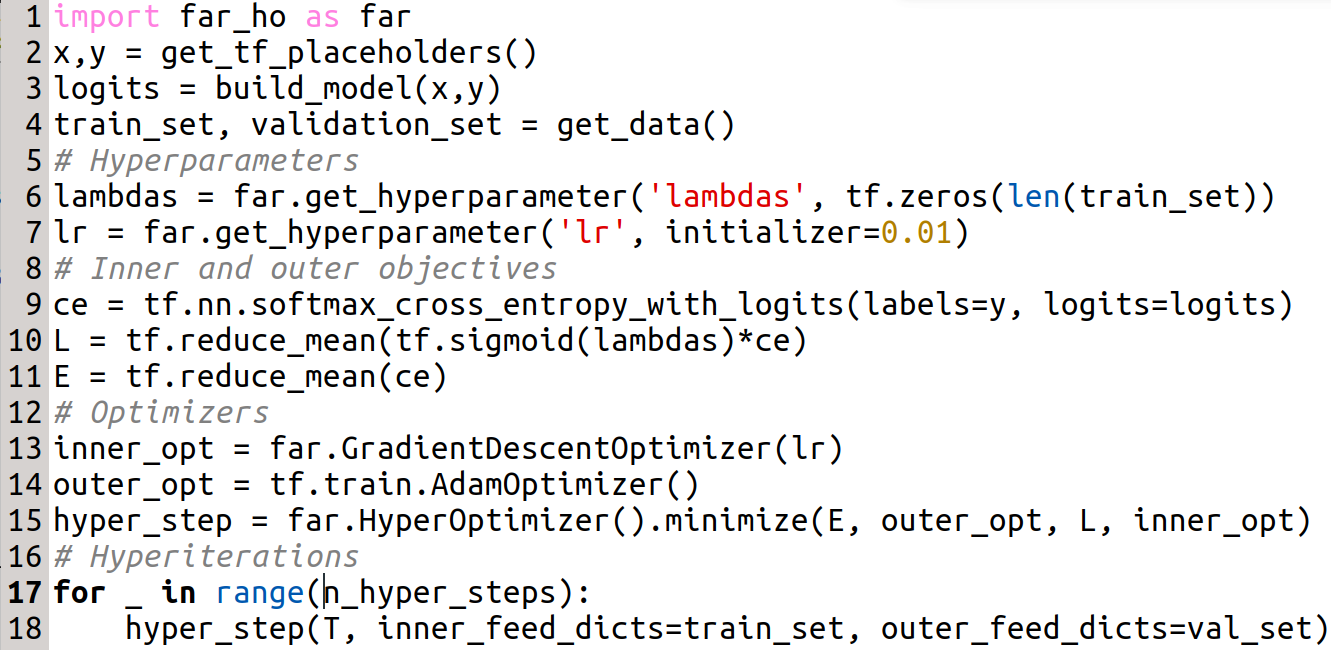}
\caption{Sample HO procedure with \textsc{Far-HO}.}\label{ex:ho}
\end{figure}

\subsection*{Hyperparameter Optimization: Weighting The Examples' Losses}
The illustrative example in Figure~\ref{ex:ho} shows how to optimize 
hyperparameters that weigh the contribution of each training example 
to the total loss. This setting is useful when
part of the training data is corrupted as in the data hyper-cleaning
experiments in~\citep{franceschi_forward_2017}.
In this example, the inner objective is $L_\lambda(w) = \sum_{(x_i,y_i)\in D_{\mathrm{tr}}} \lambda_i \ell(g_w(x_i),y_i)$ 
We treat also the learning rate as an hyperparameter (Lines 7 and 13).
As illustrated in Line 15, hyperparameters can be ``declared'' using
the \texttt{get\_hyperparameter} method, which mimics TensorFlow's
\texttt{get\_variable}, and can be placed anywhere in a computational
graph.  The method \texttt{minimize} (closely related to
\texttt{minimize} in TensorFlow) accepts two 
scalar tensors
(\texttt{E} and \texttt{L}, defined in Lines 10-11) for the outer and
the inner problem, respectively, and two associated optimizers,
\texttt{outer\_opt} and \texttt{inner\_opt} (one of which is directly
taken from TensorFlow in this example).


\begin{figure}[hb!]
\centering
\includegraphics[width=0.8\textwidth]{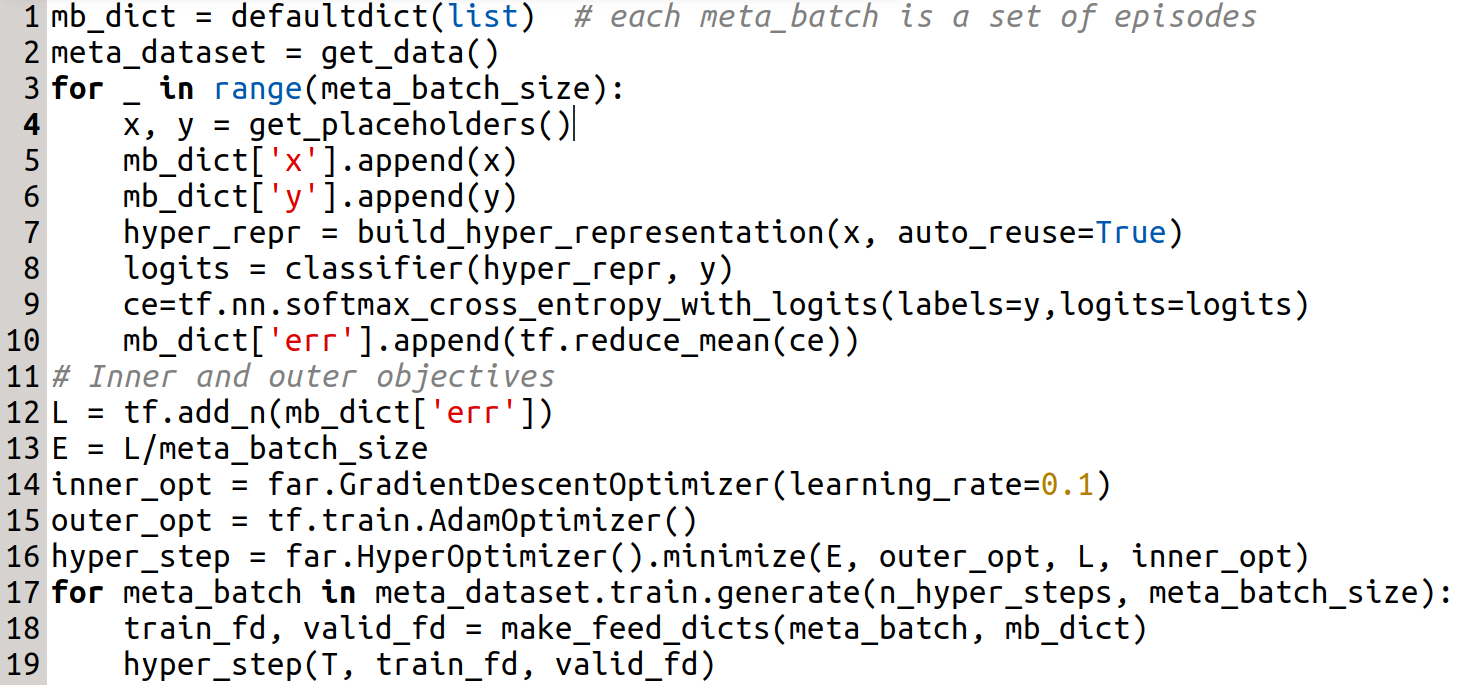}
\caption{Sample ML procedure with \textsc{Far-HO}.}\label{fig:ml}
\end{figure}

\subsection*{Meta-Learning: Hyper-Representation Networks}
Our approach \citep{franceschi_bilevel_2018} to meta-learning consists in partitioning
the model into a cross-task intermediate hyper-representation mapping
$h_\lambda:\mathcal{X}\to\mathbb{R}^k$ (parametrized by a vector $\lambda$)
and task specific models
$g^j:\mathbb{R}^k\to\mathcal{Y}^j$ (parametrized by vectors $w^j$).
The final ground model for task $j$ is thus given by $g^j\circ h$;
$\lambda$ and $w_j$ are learned by respectively optimizing the
outer and the inner objective in \eqref{eq:mlLE}.
This method is inspired by~\citep{baxter1995learning,caruana_multitask_1998}
which instead jointly optimize both hyper-representation and task specific weights.

We instantiate the approximation scheme in Problem \eqref{eq:general:constrained}
in which the weights of the task-specific models can be learned by $T$ iterations
of gradient descent, resulting in the problem:
$\min\limits_{\lambda} f_T(\lambda) = \sum_{j=1}^N L^j(w^j_T, \lambda, D^j_{\operatorname{val}})$
with $ w^j_{t} =  w^j_{t-1} - \eta \nabla_w L^j(w^j_{t-1}, \lambda, D^j_{\operatorname{tr}})
,~t, j \in [T], [N]$.
Since, in general, the number of episodes in a 
meta-training set is large, we compute a stochastic approximation of the gradient
of $f_T$ by sampling a mini-batch of episodes.  
%
Illustrative code for this example is presented in Figure~\ref{fig:ml}.


\bibliographystyle{natbib}
\bibliography{all_refs}

\begin{thebibliography}{18}
\providecommand{\natexlab}[1]{#1}
\providecommand{\url}[1]{\texttt{#1}}
\expandafter\ifx\csname urlstyle\endcsname\relax
  \providecommand{\doi}[1]{doi: #1}\else
  \providecommand{\doi}{doi: \begingroup \urlstyle{rm}\Url}\fi

\bibitem[Abadi et~al.(2015)Abadi, Agarwal, and et.
  al.]{tensorflow2015-whitepaper}
Mart\'{\i}n Abadi, Ashish Agarwal, and et. al.
\newblock {TensorFlow}: Large-scale machine learning on heterogeneous systems,
  2015.
\newblock URL \url{https://www.tensorflow.org/}.
\newblock Software available from tensorflow.org.

\bibitem[Bard(2013)]{bard_practical_2013}
Jonathan~F. Bard.
\newblock \emph{Practical bilevel optimization: algorithms and applications},
  volume~30.
\newblock Springer Science \& Business Media, 2013.
\newblock 01251.

\bibitem[Baxter(1995)]{baxter1995learning}
Jonathan Baxter.
\newblock Learning internal representations.
\newblock In \emph{Proceedings of the 8th Annual Conference on Computational
  Learning Theory ({COLT})}, pages 311--320. ACM, 1995.

\bibitem[Caruana(1998)]{caruana_multitask_1998}
Rich Caruana.
\newblock Multitask learning.
\newblock In \emph{Learning to learn}, pages 95--133. Springer, 1998.
\newblock 02683.

\bibitem[Colson et~al.(2007)Colson, Marcotte, and Savard]{colson2007overview}
Beno{\^\i}t Colson, Patrice Marcotte, and Gilles Savard.
\newblock An overview of bilevel optimization.
\newblock \emph{Annals of operations research}, 153\penalty0 (1):\penalty0
  235--256, 2007.

\bibitem[de~Freitas(2016)]{de_freitas_learning_2016}
Nando de~Freitas.
\newblock Learning to {Learn} and {Compositionality} with {Deep} {Recurrent}
  {Neural} {Networks}: {Learning} to {Learn} and {Compositionality}.
\newblock In \emph{Proceedings of the 22Nd {ACM} {SIGKDD} {International}
  {Conference} on {Knowledge} {Discovery} and {Data} {Mining}}, 2016.

\bibitem[Domke(2012)]{domke_generic_2012}
Justin Domke.
\newblock Generic {Methods} for {Optimization}-{Based} {Modeling}.
\newblock In \emph{{AISTATS}}, volume~22, pages 318--326, 2012.
\newblock URL
  \url{http://www.jmlr.org/proceedings/papers/v22/domke12/domke12.pdf}.

\bibitem[Finn et~al.(2017)Finn, Abbeel, and Levine]{finn_model-agnostic_2017}
Chelsea Finn, Pieter Abbeel, and Sergey Levine.
\newblock Model-agnostic meta-learning for fast adaptation of deep networks.
\newblock In \emph{Proceedings of the 34th International Conference on Machine
  Learning, {(ICML)}}, pages 1126--1135, 2017.
\newblock URL \url{http://proceedings.mlr.press/v70/finn17a.html}.

\bibitem[Flamary et~al.(2014)Flamary, Rakotomamonjy, and
  Gasso]{flamary2014learning}
R{\'e}mi Flamary, Alain Rakotomamonjy, and Gilles Gasso.
\newblock Learning constrained task similarities in graph-regularized
  multi-task learning.
\newblock \emph{Regularization, Optimization, Kernels, and Support Vector
  Machines}, page 103, 2014.

\bibitem[Franceschi et~al.(2017)Franceschi, Donini, Frasconi, and
  Pontil]{franceschi_forward_2017}
Luca Franceschi, Michele Donini, Paolo Frasconi, and Massimiliano Pontil.
\newblock Forward and reverse gradient-based hyperparameter optimization.
\newblock In \emph{Proceedings of the 34th International Conference on Machine
  Learning, {(ICML)}}, pages 1165--1173, 2017.
\newblock URL \url{http://proceedings.mlr.press/v70/franceschi17a.html}.

\bibitem[Franceschi et~al.(2018)Franceschi, Frasconi, Salzo, Grazzi, and
  Pontil]{franceschi_bilevel_2018}
Luca Franceschi, Paolo Frasconi, Saverio Salzo, Riccardo Grazzi, and
  Massimiliano Pontil.
\newblock Bilevel programming for hyperparameter optimization and
  meta-learning.
\newblock In \emph{Proceedings of The 35rd International Conference on Machine
  Learning {(ICML)}}, 2018.

\bibitem[Griewank and Walther(2008)]{griewank2008evaluating}
Andreas Griewank and Andrea Walther.
\newblock \emph{Evaluating derivatives: principles and techniques of
  algorithmic differentiation}.
\newblock SIAM, 2008.

\bibitem[Hutter et~al.(2015)Hutter, Lücke, and
  Schmidt-Thieme]{hutter_beyond_2015}
Frank Hutter, Jörg Lücke, and Lars Schmidt-Thieme.
\newblock Beyond {Manual} {Tuning} of {Hyperparameters}.
\newblock \emph{KI - K{\"u}nstliche Intelligenz}, 29\penalty0 (4):\penalty0
  329--337, November 2015.
\newblock ISSN 0933-1875, 1610-1987.
\newblock \doi{10.1007/s13218-015-0381-0}.
\newblock URL \url{http://link.springer.com/10.1007/s13218-015-0381-0}.

\bibitem[Keerthi et~al.(2007)Keerthi, Sindhwani, and
  Chapelle]{keerthi2007efficient}
S~Sathiya Keerthi, Vikas Sindhwani, and Olivier Chapelle.
\newblock An efficient method for gradient-based adaptation of hyperparameters
  in svm models.
\newblock In \emph{Advances in Neural Information Processing Systems {(NIPS)}},
  pages 673--680, 2007.

\bibitem[Kunapuli et~al.(2008)Kunapuli, Bennett, Hu, and
  Pang]{kunapuli_classification_2008}
G.~Kunapuli, K.P. Bennett, Jing Hu, and Jong-Shi Pang.
\newblock Classification model selection via bilevel programming.
\newblock \emph{Optimization Methods and Software}, 23\penalty0 (4):\penalty0
  475--489, August 2008.
\newblock ISSN 1055-6788, 1029-4937.
\newblock \doi{10.1080/10556780802102586}.
\newblock URL
  \url{http://www.tandfonline.com/doi/abs/10.1080/10556780802102586}.

\bibitem[Maclaurin et~al.(2015)Maclaurin, Duvenaud, and
  Adams]{maclaurin_gradient-based_2015}
Dougal Maclaurin, David~K. Duvenaud, and Ryan~P. Adams.
\newblock Gradient-based hyperparameter optimization through reversible
  learning.
\newblock In \emph{Proceedings of the 32nd International Conference on Machine
  Learning, {(ICML}}, pages 2113--2122, 2015.

\bibitem[Pedregosa(2016)]{pedregosa2016hyperparameter}
Fabian Pedregosa.
\newblock Hyperparameter optimization with approximate gradient.
\newblock In \emph{Proceedings of The 33rd International Conference on Machine
  Learning {(ICML)}}, pages 737--746, 2016.
\newblock URL \url{http://proceedings.mlr.press/v48/pedregosa16.html}.

\bibitem[Wichrowska et~al.(2017)Wichrowska, Maheswaranathan, Hoffman,
  Colmenarejo, Denil, Freitas, and Sohl-Dickstein]{wichrowska2017learnedICML}
Olga Wichrowska, Niru Maheswaranathan, Matthew~W Hoffman, Sergio~G{\'o}mez
  Colmenarejo, Misha Denil, Nando Freitas, and Jascha Sohl-Dickstein.
\newblock Learned optimizers that scale and generalize.
\newblock In \emph{Proceedings of the 34th International Conference on Machine
  Learning {(ICML)}}, pages 3751--3760, 2017.

\end{thebibliography}

\end{document}